\documentclass[3p]{elsarticle}

\usepackage{graphicx}
\usepackage{dcolumn}
\usepackage{bm}
\usepackage{epstopdf}
\usepackage[mathlines]{lineno}
\usepackage{array}
\usepackage[table]{xcolor}

\begin{document}

\title{Current frontiers in mobility and well-being}

\author[inst1]{Wen-Jing Li}
\author[inst3]{Luo-Luo Jiang \corref{correspondingauthor1}}
\author[inst4]{Attila Szolnoki \corref{correspondingauthor2}}

\address[inst1]{School of Computer Science and Technology, Zhejiang University of Water Resources and Electric Power, Hangzhou, 310018, China}
\address[inst3]{School of Information Engineering and Artificial Intelligence, Zhejiang University of Finance and Economics, Hangzhou, 310018, China}
\address[inst4]{Institute of Technical Physics and Materials Science, Centre for Energy Research, 1525 Budapest, Hungary}

\cortext[correspondingauthor1]{Corresponding author: jiangluoluo@zufe.edu.cn (L.-L. Jiang)}

\cortext[correspondingauthor2]{Corresponding author: szolnoki.attila@ek-cer.hu (A. Szolnoki)}

\date{\today}

\begin{abstract}

Human mobility has a profound impact on social structure and collective behavior, which allows individuals to reshape group structures in response to strategy information, acting as a double-edged sword for cooperative behavior. It can both disrupt existing cooperative communities and enable individuals to escape adverse environments and seek high-quality interaction partners. This paper surveys current research progress on the impact of mobility on cooperation within the framework of evolutionary game theory, focusing on specific mechanisms of passive escape, dynamic self-organization driven by active choice, and the core trade-off between migration costs and speed. Mobility, an individual's spatial response to local strategic information that turns location into strategic advantage, influences cooperative evolution by altering interactive information and environmental feedback, consequently affecting sustainable social well-being, such as collective benefits. Under specific conditions, this self-interested behavior at the individual level can foster macrosocial harmony, embodying the unity of individual rationality and collective cooperation. Furthermore, we also discusses inter-layer migration and information routing mechanisms in multilayer networks and extends the perspective to human-machine hybrid systems where mobile agents with information regulation capabilities can serve as information supernodes. Our paper aims to provide a unified framework for understanding mobility as a combined physical and informational driving force in the evolution of cooperation.

\end{abstract}

\maketitle

\section{\label{sec:level1}Introduction}
\indent From early human ancestors leaving Africa to contemporary cross-border migration, human movement has permanently been shaping our world. Beyond its macro-level significance, individual mobility exhibits universal statistical patterns and forms the micro-foundation of social functioning~\cite{Brockmann_Nature2006,Gonzalez_Nature2008,Alessandretti_Nature2020}. Mobility not only profoundly impacts individual lives but also brings a significant influence on social structures and collective behavior of human beings~\cite{Barbosa_PhysRep2018}. The concept of mobility were already embedded in the wisdom of human civilization long before the birth of modern science. The ancient Chinese story of Meng Mu's Three Moves  serves as an excellent example. More than two thousand years ago, the mother of the philosopher Mencius moved her family three times in order to provide her son with a favorable environment in which to grow up. It was precisely these repeated moves that profoundly shaped Mencius's development, demonstrating just how crucial adapting to environmental changes is to human behavior. The essence of movement lies in the fact that it is an active or passive choice made by the human beings in relation to the spatial environment. The three relocations were, in essence, an effort to choose a community for her family that was more conducive to cooperation and learning. This ancient wisdom resonates with modern exploration of the complex relationship between mobility and cooperative behavior.

Among the many mechanisms of cooperation~\cite{Nowak2006,perc_pr17,xie_k_amc26,liu_lj_el23}, including direct reciprocity, indirect reciprocity, kin selection, and punishment, spatial structure has long been regarded as a key mechanism promoting the emergence of cooperation. Individuals in a spatial context can increase the probability of interaction by forming protective clusters, thereby sustaining cooperation, which is known as network reciprocity~\cite{SzolnokiEPL2007,PercPRE2008}. In spatial network reciprocity, however, the introduction of individual mobility makes the picture more subtle and yields seemingly contradictory effects. Whereas mobility may disrupt established cooperative clusters, exposing cooperators to exploitation by cheaters, it also offers cooperators the opportunity to escape unfavorable environments~\cite{XiaoNJP2020,zhao_xw_csf26}. In this sense, mobility can be viewed as a mechanism for local mixing, where greater mobility makes it easier for the advantages that spatial structure originally provided for cooperators to be undermined~\cite{DuhPRE2020}. Analysis of large scale mobility data has revealed a series of remarkable statistical patterns~\cite{songnatphys2010,BoucherieNatHumBehav2025}. It was found that the distribution of human travel distances exhibits significant power-law characteristics, indicating that while most people have limited travel ranges, a small number of individuals engage in long-distance travel~\cite{SongScience2010}. This heterogeneity profoundly influences the processes of spatial interaction and the propagation of cooperative behavior~\cite{MeloniPRE2009}. The key lies in how mobility interacts with strategies and environmental feedback, thereby giving rise to processes of passive escape and dynamic self-organization~\cite{HelbingACS2008,JiangPRE2010,SicardiJTB2009}.

While passive escape merely allows individuals to evade unfavorable conditions, dynamic self-organization is largely driven by active selection. An individual's migration to a high-yield environment often facilitates the replication and adoption of a successful strategy~\cite{HelbingPNAS2009}. When this choice is embedded within a spatial network, it creates a positive feedback mechanism that drives cooperative behavior to expand from local advantages to global order~\cite{HelbingACS2008,JiangPRE2010}. Unlike the movement in passive flight, which lacks a clear destination, under the active screening mechanism, individuals can proactively evaluate and select migration destinations based on specific criteria (such as expected returns) to maximize their own interests~\cite{WuPRE2012, ChengNJP2011}. At its core lies the tendency of rational individuals to vote with their feet, which is migrating to environments that offer higher returns or high-reputation partners~\cite{CongPLoSONE2012,LiAMC2018}. This active selection process creates a powerful positive feedback loop in the evolution of cooperation, where individuals actively migrate in search of more favorable interaction environments or high-reputation partners, which promotes the spatial clustering of cooperators~\cite{CardinotNJP2019, IchinoseEPL2018}. The expansion and stability of these clusters, in turn, enhance the fitness of cooperators, attracting more individuals to join the cooperative camp~\cite{GelimsonPRE2013, AmorPRE2011}. However, whether movement is passive or active, it is inevitably subject to real-world constraints. The resilience of the broader cooperative order thus depends on how these complex micro-level choices are made under such limitations, including robust social structures, and operational efficiency~\cite{SicardiJTB2009,VainsteinJTB2007,IchinoseEM2008,AktipisEHB2011, LinCSF2011}.

Mobility is not simply an external perturbation in the evolutionary process, but could form a bidirectional feedback with strategy dynamics~\cite{PercBioSystems2010}. The movement of players alters the interactive environment and information feedback, which in turn influences individual movements and strategic choices. In real-world scenarios, every migration requires individuals to carefully balance migration costs against migration speed~\cite{LiuCSF2012,ChenPRE2012}. Within the framework of spatial evolutionary games, individual movement is no longer purely a simple matter of seeking benefits and avoiding harm, but rather a complex decision-making process involving cost assessment, speed regulation, and information responsiveness~\cite{GelimsonPRE2013, ChenPA2011,YangPRE2010}. Early research often assumed that movement was cost-free or focused solely on its role in promoting cooperation~\cite{IchinoseSR2013, TomassiniJTB2015}. However, migration in real-world is frequently accompanied by multiple costs, including information acquisition and environmental adaptation~\cite{CongSR2017}. In recent years, an increasing number of studies have begun to systematically examine the independent and interactive effects of these two key dimensions, such as migration cost and migration speed~\cite{AntonioniSR2014,FunkPC2019,LiAMC2020, LiWamc2022}. These studies on the trade-off between cost and speed reveal that faster movement does not necessarily promote cooperation, and lower costs are not always the optimal choice~\cite{LiPA2019, WangJTB2015}. The optimal migration strategy requires striking a delicate balance between individual payoffs, influenced by cost and speed, and spatial structure~\cite{SadhukhanPR2021}. Cost stabilizes cooperative clusters by screening migrants and regulating migration frequency, asymmetrically weakening defectors~\cite{CongPLoSONE2012,ArmanoSR2017}. Intriguingly, in certain scenarios, the rapid movement of defectors may paradoxically enhance global stability~\cite{GelimsonPRE2013,YoungTBP2018,LiCSF2023}. These spreading processes of cooperation reveal dual roles and nonlinear influences of mobility~\cite{ZhaoCSF2023,LiNonD2024}. In short, studying how mobility influences social cooperation is not only of profound theoretical significance but also has broad practical applications across multiple disciplinary fields. For example, in the currently booming fields of artificial intelligence and robotics, the design of collaborative robots must be based on a deep understanding of the effect of mobility on cooperative behavior~\cite{Spnasn2024,SharmaJRoySocInterface2023,SiJRoySocInterface2025}. How is cooperation achieved within a team of mobile robots? How should mobility strategies be designed to effectively promote collective cooperation? The answers to these questions depend on systematic research into migration patterns, the dynamics of spatial structures, and the evolution of strategies. This review of spatial mobile games provides a foundational framework and forward-looking theoretical guidance for addressing these questions. The ultimate social goal of human mobility and cooperation is human well-being, which means the macro-level outcomes of quality of life, social resilience, and collective benefits at both the individual and group levels. This is because cooperative humans form the foundation for society to achieve collective benefits and social cohesion. Consequently, the evolution of cooperation can be viewed as one of the key pathways to achieving both individual and collective well-being. Building on this premise, we focus on the evolution of cooperation as an intermediate mechanism of fundamental importance to well-being. In spatial evolutionary games, mobility influences the emergence and maintenance of cooperative behavior by altering the feedback loops involving interaction partners, the information, and the environment. Changes in the level of cooperation, in turn, affect collective benefits, social resilience, and public goods at the group level, which is precisely what we define as sustainable well-being.

This paper adopts the format of a narrative review, focusing on studies that reflect key developments and representative achievements in this field. For literature search, we primarily relied on databases such as Web of Science, Scopus, and Google Scholar, conducting search queries on topics of mobile games and complex networks. The time frame focused on the past two decades approximately, while also incorporating seminal early works in the field to illustrate the continuity and historical context of the development of mobile games in complex networks. In screening representative studies, we first assessed whether the studies were directly relevant to network structure, game dynamics under mobility, and cooperative evolution mechanisms. Second, we examined their theoretical or empirical contributions, including whether they proposed important models, mechanisms, analytical methods, or experimental findings, as well as the impact and peer recognition of the journals in which they were published. Finally, we considered citation patterns and subsequent influence, giving priority to important literature that has spurred follow-up research. Our paper is structured as follows: First, we discuss relevant works on mobility with passive escape and dynamic self-organization in Sec.~2. Next,
in Sec.~3, we discuss proactive screening and partner selection on mobility. Then, in Sec.~4, we look into evaluating the advantages and disadvantages to move. Most importantly, we review progress on spreading of information and strategy in Sec.~5. Finally, future applications in AI-Human hybrid systems and discussion are provided in Sec.~6 and Sec.~7, respectively.

\begin{figure*}
\centering
\scalebox{0.7}[0.7]{\includegraphics{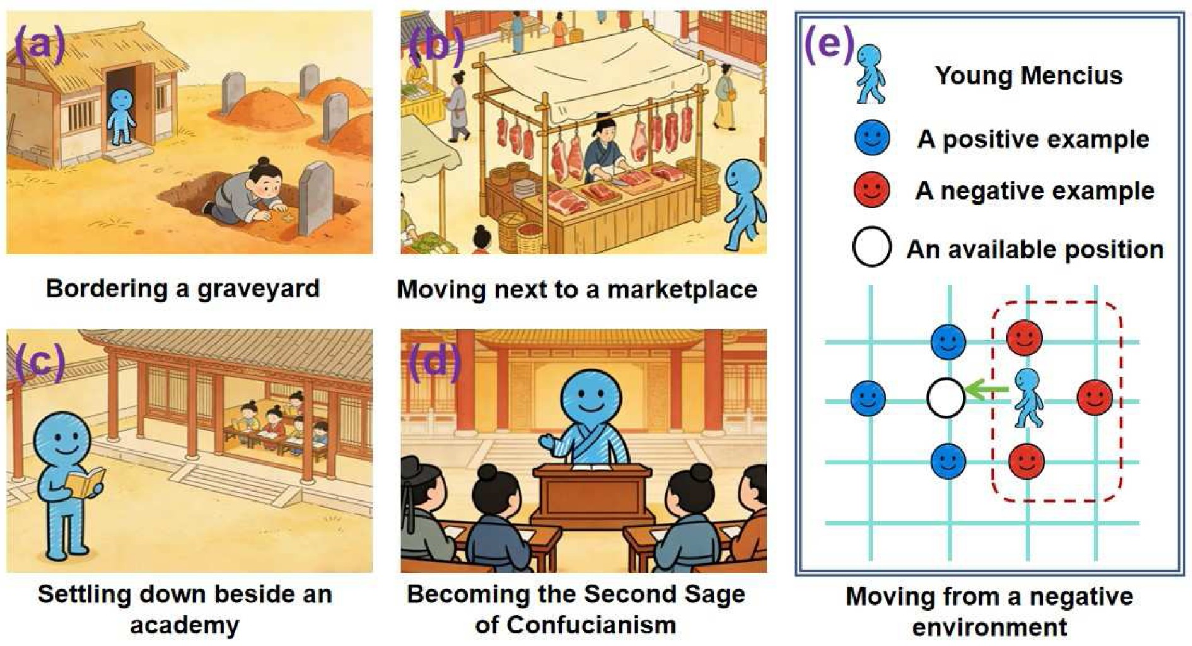}}
\caption{Illustration of Three Moves of the Mencius' Mother and the concept of moving from a negative environment. Further explanation can be found in the main text.
\label{Fig:fig1} }
\end{figure*}

\section{Passive escape and dynamic self-organization}\label{sec:Passive}
The ancient story of Meng Mu's Three Moves illustrates how people often relocate out of necessity to change their current circumstances. Just as in real life, where many people move due to economic pressures, social exclusion, their migration is often a passive response to external pressures. As shown in Fig.~\ref{Fig:fig1}, during the Warring States period in ancient China, the mother of Mencius, the Confucian philosopher known as the Second Sage, moved three times to provide him a better learning environment. Their first home was near a cemetery. Young Mencius played there, but soon began imitating the funeral rituals he saw, pretending to weep and kneel like the others. Seeing this, Mencius' mother deemed it unsuitable. They then relocated to a bustling marketplace, where crowds thronged and vendors sold meat. In this second neighborhood, young Mencius began imitating the butcher next door. Once more, Mencius' mother moved the family, this time settling beside a schoolhouse. Young Mencius would holding bamboo slips, earnestly reciting along with the voices drifting from an academy. Thus, Mencius ultimately became a world-renowned scholar. This story faithfully illustrates the learning principle of associate with the virtuous, distance yourself from the wicked, which profoundly influences future generations. As noted, spatial structure has long been recognized as a key mechanism for fostering cooperation, enabling collaborators to form and maintain clusters that increase the probability of interaction. When individual spatial mobility is introduced, its effects become complex and seemingly contradictory: on one hand, movement may disrupt collaborative clusters, exposing partners to exploitation by defectors; on the other hand, it may offer collaborators opportunities to escape unfavorable environments and seek or form more advantageous groups~\cite{HelbingACS2008}. Thus arises the fundamental question: What are the major motivations for migration? We summarize three core mechanisms, namely, passive escape and dynamic self-organization, proactive screening and partner selection, evaluating the potential advantages and disadvantages. To support our argument, the major research works can be classed into these categories successfully, as shown in Table~\ref{tab1}.

\begin{table}[ht]
	\centering
	\begin{tabular}{|p{2.5cm}|p{2.8cm}|p{2.8cm}|}
		\hline
		\rowcolor{gray!30}
		  & \cellcolor{yellow!20} The Prisoner's Dilemma & \cellcolor{yellow!20} Public Goods Games \\
		\hline
		\cellcolor{blue!20} Passive escape and dynamic self-organization & \cellcolor{green!20} \cite{HelbingACS2008,JiangPRE2010,CardinotNJP2019,GelimsonPRE2013,AmorPRE2011,VainsteinJTB2007,AntonioniSR2015,inabacsf2026,qiangem2025,kroumipre2025a} & \cellcolor{green!20} \cite{RocaPNAS2011,dhakalamc2020,dhakalrsos2022,jisr2025,kroumipre2025b}\\
        \hline
		\cellcolor{blue!20} Proactive screening and partner selection & \cellcolor{pink!20} \cite{HelbingPNAS2009,ChengNJP2011,CongPLoSONE2012,ChenPA2016,cheng_y_amc26b,ZhangPA2022,LiWcsf2022,wangcpb2025} & \cellcolor{pink!20} \cite{ArmanoSR2017,SiAMC2025,zhukbs2024}\\
        \hline
		\cellcolor{blue!20} Evaluating the advantages and disadvantages & \cellcolor{green!20} \cite{ZhaoCSF2023,LiNonD2024,zhangcsf2021,zhangcsf2023} & \cellcolor{green!20}\cite{RanCSF2023,GrossCP2025,xiaoepjb2022,zhangepl2021} \\
         \hline
	\end{tabular}
	\caption{Recent research papers considering mobile populations in social dilemmas through the eye of main motivations of individual movement.}
	\label{tab1}
\end{table}

\begin{figure*}[h!]
	\centering
	\scalebox{0.6}[0.5]{\includegraphics{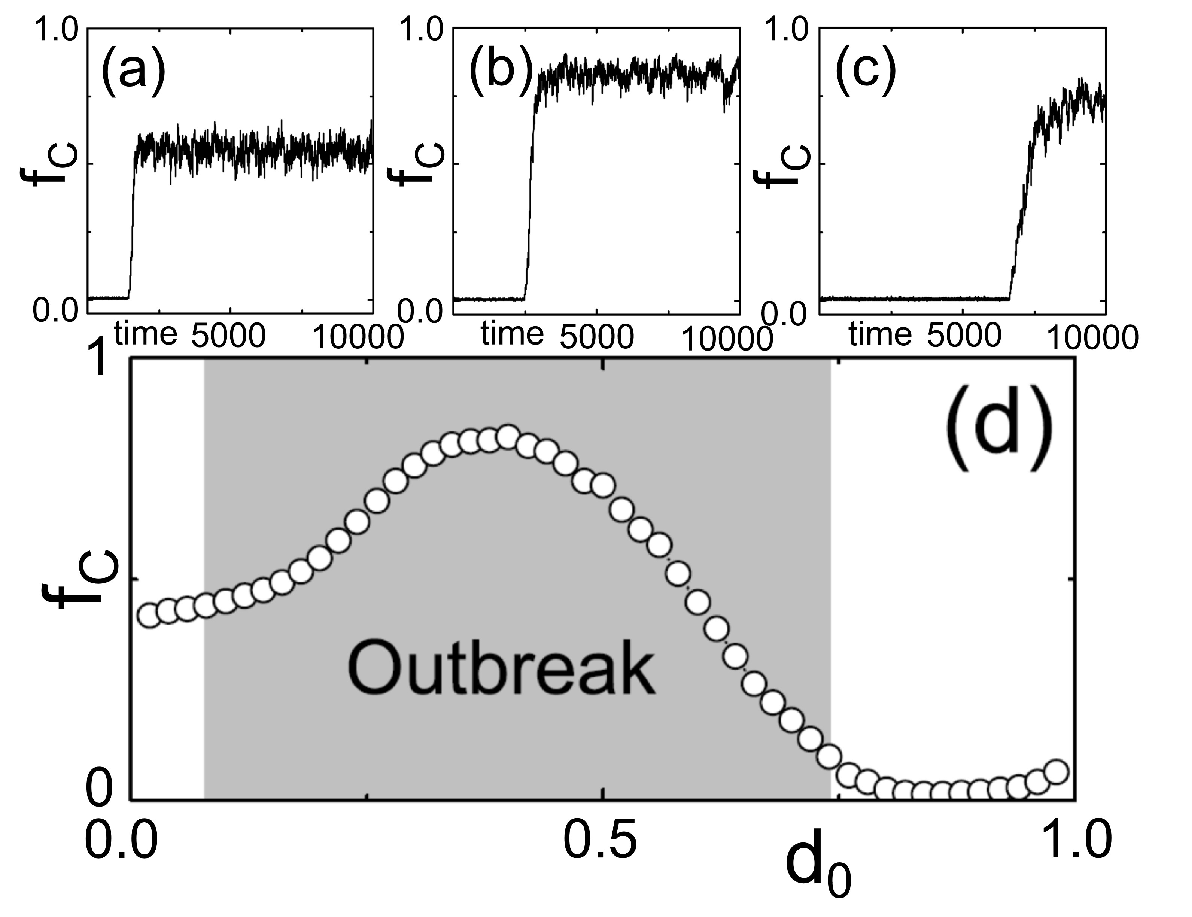}}
	\caption{The impact of empty sites on the evolution of cooperation in PDG. Panel~(d) summarizes the stationary cooperation level in dependence of the fraction of empty sites, $d_0$~\cite{JiangPRE2010}. Panels~(a)-(c) illustrates the time evolution for (a) $d_{0}=0.1$, (b) $d_{0}=0.4$, and (c) $d_{0}=0.6$. Importantly, there is a typical waiting time before the outbreak of cooperation, and the gray area in panel (d) represents the region where the outbreak of cooperation occurs. In this Prisoner's Dilemma game, the payoff obtained by players are as follows: when a cooperator encounters a cooperator, both receive a payoff of 1; when a defector encounters a cooperator, the defector receives a temptation payoff of $b=1.4$, while the cooperator receives a sucker's payoff of 0; when two defectors encounter each other, both are punished and receive a payoff of 0. The migration probability is determined by the local environment, defined as follow: For a selected individual, the migration probability is defined as the ratio of the number of defectors $n_{D}$ in its neighborhood to the node degree $k$ (e.g. $n_{D}/k$). Thus, individuals surrounded by more defectors are more likely to relocate to an empty neighboring site.
		\label{Fig:fig2-waitingtime} }
\end{figure*}

The core of passive escape lies in individuals responding on locally observable information only~\cite{JiangPRE2010}. The adaptive migration model assumes that the probability of migration depends on the number of defectors among its neighbors. In this scenario, where escaping defection is central, the evolution of cooperation requires neither foresight on the part of individuals nor complex social cognition. Simple migration rules that respond solely to immediate local risks are sufficient to build stable cooperative structures. As Fig.~\ref{Fig:fig2-waitingtime} illustrates, despite being initiated entirely by defectors, the system exhibited a prolonged latent period during which individuals evaluate local risk according to neighboring defector counts and relocates to evade areas where defectors are present with high-density density. This attitude of defection avoidance allows cooperators to break free from being surrounded and spontaneously aggregate into protective clusters. More importantly, this aggregation acts as a dynamic defensive mechanism. When defectors attempt to invade, cooperators constantly relocate and regroup in available vacancies, preventing the clusters from internal erosion. Through this kind of spatial self-organization, collective cooperation suddenly surges at the critical point, rapidly propelling the system into a state dominated by cooperators, just as a spark can ignite a prairie fire. This sudden outbreak demonstrates the robust resilience of cooperative dynamics. These simple rules offer the following advantages:
(i) Using locally available information: Individuals need only know the strategy of their immediate neighbors (cooperators or defectors). This information is naturally obtained through the previous round of game interactions, requiring no additional scouting or communication costs.
(ii) Aligned incentives: Regardless of whether an individual is a cooperator or defector, interacting with cooperators yields higher returns than interacting with defectors. Thus, escaping environments dense with defectors and seeking out environments rich in cooperators becomes a universal, self-interested driving force. For cooperators, this avoids exploitation; for defectors, it seeks more targets to exploit.
(iii) Achieving Passive Segregation: This rule inherently drives  aversion-based  movement. Areas with high defector density trigger higher out-migration rates, while cooperator-dense zones remain more stable. This gradually leads to spatial self-organization: Cooperators cluster together to form relatively pure groups, while defectors are pushed to the periphery of these clusters or into isolated patches.

The above discussed analysis has thoroughly examined how the simple rule of passive avoidance serves as the spark for cooperation to spontaneously emerge within spatial structures. The next question is how to transform this naive logic, based on local reactions, into a testable model capable of describing more complex cooperative evolutionary processes. This leads us to the research tradition of evolutionary games with mobile players. Pioneering early work laid the foundation for understanding passive escape mechanisms. Unbiased always-move rules in the Prisoner's Dilemma could enhance cooperation compared to static models, even enabling cooperation to emerge from a population of defectors~\cite{VainsteinJTB2007}. Subsequently, cooperation could persist among two-dimensional moving Prisoner's Dilemma players, provided the temptation to defect and movement speed remained moderate, with the system gradually converging toward a fully cooperative state~\cite{MeloniPRE2009}. Furthermore, the classic walk away rule was first proposed, demonstrating that individuals need only leave low-payoff groups to significantly promote the evolution of cooperation, with cooperators gradually aggregating into stable large groups~\cite{AktipisEHB2011}. This reveals the powerful force of the simple rule of escaping unfavorable environments. When random movement is introduced into the spatial N-player iterative Prisoner's Dilemma, it accelerates the population's convergence to equilibrium, highlighting mobility's role in accelerating spatial reorganization~\cite{ChiongPA2012}. Even without active partner selection, simple movement rules can trigger self-segregation by altering spatial connections between individuals. This occurs because mobility creates cohesion differences between cooperators and defectors, enabling cooperators to form tighter clusters through movement, thereby achieving positive selection sorting~\cite{JoshiPCB2017}. This dynamic perspective on self-segregation elevates mobility from mere escape to a mechanism for group formation.

Random movement can indeed sometimes promote cooperation, but this is largely a side effect of changes in the probability of encounters, and its impact is limited~\cite{SicardiJTB2009}. The more fundamental driving force is enabling individuals to escape from cheaters under certain conditions. The truly potent driving force stems from individuals' responses to environmental signals. Crucially, a golden rule for directed migration was identified: Near the virtuous, far from the wicked, which allows cooperators to escape defectors and find each other, forming cooperation islands in an early theoretical exposition of how passive escape promotes spatial self-organization~\cite{HelbingACS2008}. Even without active choice, simple movement rules can foster cooperation through spatial strategy correlations~\cite{JoshiPCB2017}. However, conditionally triggered movements exhibit greater ecological relevance and power. Unconditional L{\'e}vy flights hinder cooperation, but when jumps respond to increased betrayal among neighbors, they significantly boost cooperation in various games, including the Prisoner's Dilemma~\cite{TomassiniJTB2015}. The pivotal role of responsive movement is first evident in  escape  preceding  confrontation. When facing environmental risks, an individual's first instinct is often not to stay and actively change, but to migrate and avoid being exploited. This explains why community decline is often difficult to reverse. Second, an individual's decision basically relies on available local signals. Thus, effective management hinges on influencing these visible cues rather than relying solely on broad messaging. Finally, systems possess a risk perception threshold~\cite{wang_l_pcbi26}. Once crossed, it triggers mass migration waves, leading to structural change.

\begin{figure*}
\centering
\scalebox{0.4}[0.4]{\includegraphics{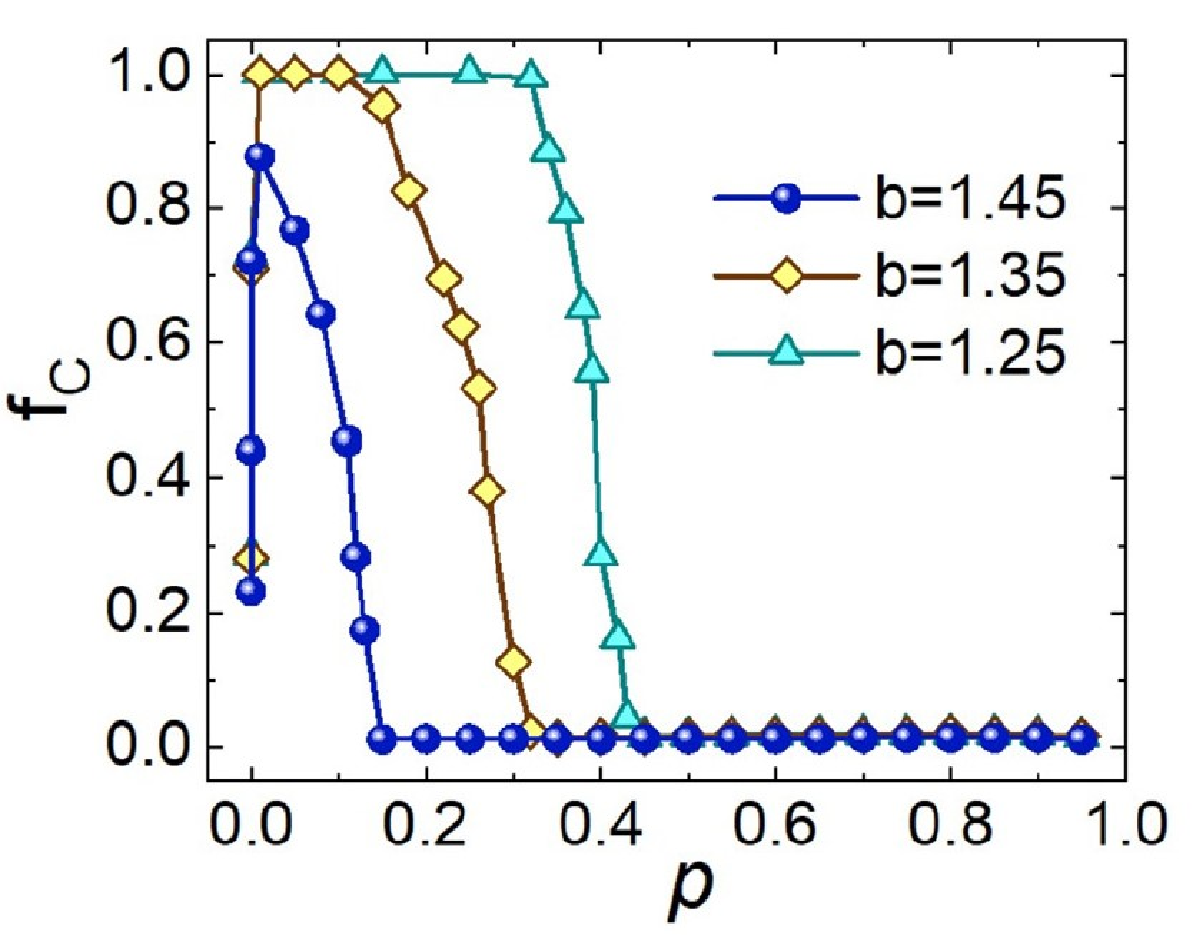}}
\caption{The curve of cooperation frequency $f_{C}$ with $p$~\cite{LiWcsf2022}, where the parameter $p$ is defined as follows: a low-income individual either moves into an empty site with probability $p$, provided at least one empty site available in its direct neighborhood, or engages in social learning to update its strategy with probability $1- p$. In this Prisoner's Dilemma game, the payoff obtained by players are as follows: when a cooperator encounters a cooperator, both receive a payoff of 1; when a defector encounters a cooperator, the defector receives a temptation payoff of $b$, while the cooperator receives a sucker's payoff of -0.1; when two defectors encounter each other, both are punished and receive a payoff of 0. Furthermore, the empty sites density is set as 0.3, and the system consists of a $128\times128$ lattice.
\label{Fig:fig3-lowincome} }
\end{figure*}

From undirected random walks to conditional migration based on reward feedback, this cognitive leap reveals the underlying logic of social self-organization. Mobility is not only a means for individuals to pursue self-interest, but also redefines group structures of society on a macro scale. Even in societies composed of greedy, mobile individuals with the ``win-stay, lose-move'' can spontaneously give rise to high social cohesion through the co-evolution of cooperation and spatial organization~\cite{RocaPNAS2011}. In voluntary prisoner's dilemmas, allowing agents to relocate based on success can automatically repair cooperative mechanisms broken by purely rational decision updates~\cite{CardinotNJP2019}. Migration that considers overall social welfare can also promote cooperation and generate distinctive spatial patterns~\cite{LiPA2016}. The migration of low-income groups is a specific example to support this logic. Given a fixed vacancy density of 0.3, Fig.~\ref{Fig:fig3-lowincome} shows how the cooperation frequency $f_{C}$ evolves as a function of the probability of migration $p$ for low-income individuals under different incentive factors $b$ (Here, if a player had a lower payoff than the average payoff in his/her community, then the player was classified as a low-income one). The three curves exhibit consistent non-monotonic behavior. When $p = 0.00$, the system completely prohibits migration, and cooperation is sustained solely by network reciprocity, with $f_{C}$ approximately equal to 0.20. However, when $p$ increased slightly from zero, $f_{C}$ undergoes a dramatic jump and peaks near $p = 0.01$. Taking $b = 1.45$ as an example, the cooperation rate $f_{C}$ jumps abruptly from approximately 0.23 to about 0.88. This counterintuitive result suggests that even an extremely low level of migration is sufficient to trigger a qualitative change in the level of cooperation. Moderate migration among low-income individuals can significantly enhance the level of social cooperation through two pathways: preserving the framework of network reciprocity and activating the competitive advantages of migration. However, once the migration frequency exceeds a certain threshold, the losses resulting from the breakdown of network structure will outweigh the benefits derived from the migration mechanism itself, and cooperation will actually be undermined.

Mobility often integrates with other social response mechanisms to form richer cooperative maintenance systems. When individuals can not only escape but also employ complementary strategies like punishment and risk assessment, the foundation of cooperation becomes more robust. Even minor diffusive mobility severely constrains cooperation by enabling defectors to infiltrate cooperator clusters~\cite{GelimsonPRE2013}. This highlights that the spatial prisoner's dilemma alone cannot explain stable cooperation promoted by the network reciprocity and additional conditional mechanisms on mobility are required. Experimental studies provide crucial evidence that spatial mobility had no significant impact on cooperation levels. Cooperative clusters briefly formed but rapidly disintegrated at boundaries exploited by defectors~\cite{AntonioniSR2015}. This observation contradicts many theoretical simulations, underscoring the complexity of human decision-making and the limitations of mobility. When altruistic punishment was introduced, penalizing only a minority of defectors sufficed to push the system toward cooperation~\cite{AmorPRE2011}. In risk-driven migration, players autonomously relocate when collective failure risk becomes excessive. This self-organizing mechanism promotes cooperation more effectively than manually setting migration rates and generates unexpected spatial patterns~\cite{ChenPRE2012}. The movement of punishers may enhance their survival prospects~\cite{CongSR2017}. Migration driven by the pursuit of potential gains promotes population clustering, and this pattern of spatial self-organization is crucial for the emergence of altruistic punishment. Conversely, individuals who tend to avoid losses accelerate the disintegration of cooperative clusters~\cite{GuoPA2023}.

In summary, despite the risk of disrupting cooperative clusters, spatial mobility has emerged as an effective mechanism for fostering cooperation. The golden rule could be summed as upgrading from random movement to a directional, conditional, and responsive movement based on local feedback. Such movement not only helps individuals passively escape exploitation but, more crucially, triggers spatial self-segregation and dynamic self-organization among cooperators and defectors. This, in turn, shapes a macro-level social-geographical structure conducive to cooperation.

\section{Proactive screening and partner selection}\label{sec:Proactive}

To understand how individual behavior aggregates into regular group patterns, it is essential to examine the transition from individual movement to collective flow patterns~\cite{Barbosa_PhysRep2018}. With the rapid advancement of technology, the scale of individual movement has grown increasingly large, giving rise to mass migration. The motivations and patterns underlying these mass migrations undoubtedly have a profound impact on group behavior~\cite{schlapfernature2021}. A transition from individual to collective mobility, if confined solely to the aggregation of individuals in physical space, is insufficient to reveal the underlying mechanisms of emergent collective behavior. What truly drives the formation of macroscopic order is often the proactive choices made by individuals based on their expected benefits, which is precisely the key to understanding how cooperation evolves spontaneously and efficiently within spatial structures.

In spatial structures, both passive and active behaviors of individuals can promote the spontaneous formation of cooperation. One is the passive phenomenon, such as birds of a feather flock together, where individuals simply gather together. The other is active selection with choosing the best tree to nest in. The latter is more efficient because individuals do not fly aimlessly but instead focus on their goals and select locations based on specific criteria. Within this active selection framework, individuals first consider the potential for higher returns through migration, naturally weighing expected benefits when deciding whether to move to potentially lucrative areas. Success-driven migration involves testing the potential returns of all vacant positions through fictitious play and selecting the location with the highest expected return. The position of a more successful neighbor~\cite{HelbingPNAS2009}. The advantage of this migration strategy is that when defectors emerge within cooperative clusters, their high returns attract imitation, temporarily spreading defection. However, success-driven migration provides cooperators within clusters an escape route from defector exploitation. By reconfiguring positions through migration, cooperators may trigger the fragmentation of large clusters, ultimately marginalizing defectors to the periphery of newly formed smaller clusters. At the boundary, defectors often yield lower returns than individuals within the cluster who have more cooperating neighbors. Consequently, defectors are more likely to switch to cooperation during subsequent strategy imitation. If success-driven migration characterizes the fundamental form of active selection from the single dimension of individual payoff expectations, then further model refinement reveals that the heterogeneity of migration decisions can become a key lever for optimizing cooperative dynamics~\cite{chen_yq_csf26}. This decision heterogeneity manifests as individuals adjusting their migration propensity based on their immediate state (such as payoff levels). At high player densities, payoff-driven migration significantly enhances cooperation levels, even achieving full cooperation across a broad range of parameters. The mechanism is that high-payoff defectors migrate faster, enabling them to leave cooperative clusters more readily and reducing their erosion of the cluster, while cooperators migrate relatively slower, promoting cluster stability. Simultaneously, this heterogeneous migration alters the growth patterns of clusters, promoting the expansion of cooperative clusters~\cite{ChengNJP2011}. This benefit disparity can be further refined by linking individual migration speed to personal payoff, which proves more effective at promoting cooperation than uniform migration speeds~\cite{ChenPA2016}. Individuals in migration decisions consider not only personal gains but also potential opportunities, balancing these against migration costs  simulating rational decision-making in reality. When migration costs are low, successful migration drives selection toward advantageous locations; when costs become prohibitively high, migration is suppressed, yet cooperation can still be maintained through alternative mechanisms~\cite{LiuCSF2012}.

\begin{figure*}
\centering
\scalebox{0.6}[0.6]{\includegraphics{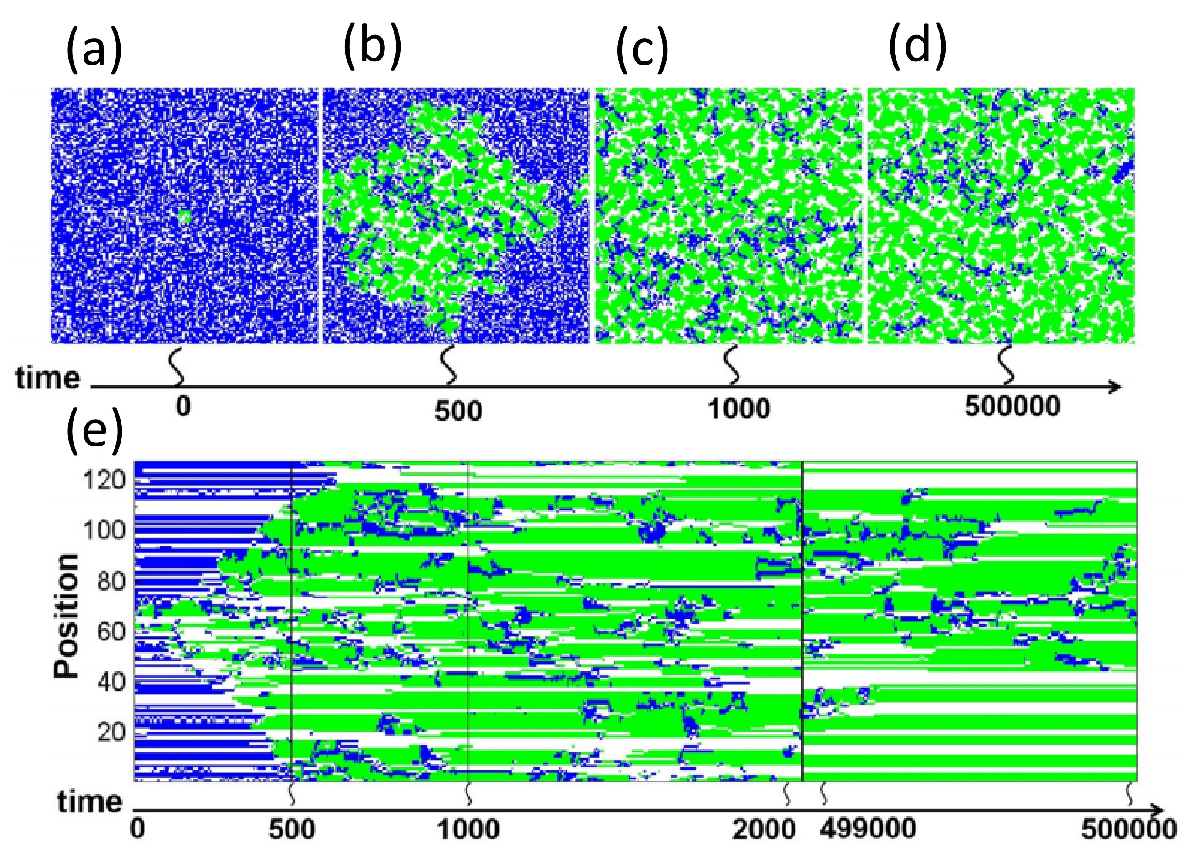}}
\caption{The representative evolutionary patterns with a bunch of cooperators besieged by all defectors~\cite{LiCSF2020}. Evolutionary process of all row sites in middle column over time (e) and the evolutionary patterns at time steps $0$ (a), $500$ (b), $1000$ (c) and $500000$ (d). The related value of cooperation level $f_{C}$ is about 0.002, 0.359, 0.818 and 0.884 respectively. Blue represents defectors, green cooperators, and white empty sites. In this Prisoner's Dilemma game, the payoff obtained by players are as follows: when a cooperator encounters a cooperator, both receive a payoff of 1; when a defector encounters a cooperator, the defector receives a temptation payoff of $b=1.5$, while the cooperator receives a sucker's payoff of 0; when two defectors encounter each other, both are punished and receive a payoff of 0. The system consists of a $128\times128$ lattice, where the portion of empty sites is 0.3.
\label{Fig:fig4-pattern} }
\end{figure*}

Unlike ordinary active migration driven by success, advantageous individuals balance their own interests with those of the community as a whole when moving. To gain a deeper understanding of the key influence that advantageous individuals have on evolutionary processes, it is necessary to further investigate models that integrate advantageous individuals with traditional models of success-driven migration, where players will move with probability $0.1$ if the action could bring more payoff to the potential community. A more challenging initial configuration positioned a sparse group of cooperators at the grid center, surrounded by multiple tiers of defectors. The analysis of the evolutionary process unfolds primarily along two dimensions, as shown in Fig.~\ref{Fig:fig4-pattern}. Firstly, observing overall snapshots of the system at key time points from panel~(a) to panel~(d). Secondly, tracking the dynamic evolution of a specific vertical cross-section in column 64 over time, illustrated by splicing together representative evolutionary segments in panel~(e). In this seemingly hopeless scenario, benefit-seeking individuals play a pivotal role in turning the tide. Unlike ordinary individuals, who focus solely on personal gain or simply seek to avoid harm, benefit-seeking individuals adopt an altruistic migration. When moving, they not only consider their own interests but also assess whether their migration will increase the average benefit of community. This is the paradigm that shifts from passive escape to active profit-seeking and triggers a powerful positive feedback loop. Individuals actively migrate to seek a more favorable interactive environment, which drives cooperators to cluster spatially, forming tight and stable interaction clusters. As a result, the expansion and stability of these clusters further enhance the fitness of cooperators, attracting more individuals to join the cooperative camp. By occupying positions that generate higher community payoffs, they not only achieve a dual increase in both individual and collective payoffs but also fundamentally reverse the inevitable outcome of cooperative extinction, ultimately leading the system into a cooperative-dominated dynamic equilibrium. This fully demonstrates that, in games fraught with uncertainty, taking the initiative often yields a greater survival advantage.

\begin{figure*}
\centering
\scalebox{0.7}[0.7]{\includegraphics{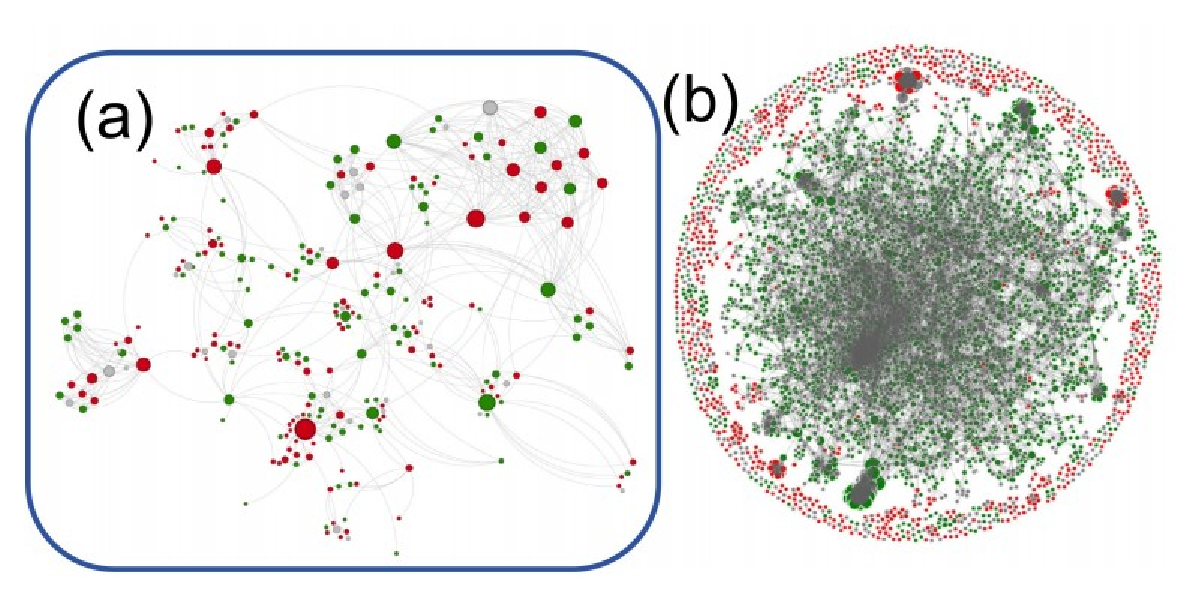}}
\caption{Cooperators survive in big groups, whereas leave sites in small groups for defectors~\cite{LiCSF2023}. Panel~(a) shows a portion of the network in its initial state (only a portion is shown for clarity, cooperators and defectors are initially distributed uniformly). Panel~(b) illustrates the pattern of the entire network as the system evolves into a stable state. Green denotes cooperators, red represents defectors, and grey spot marks empty sites. The size of the spots characterizes the degree of the site. The payoff parametrization of the Prisoner's Dilemma are as follows: players for mutual cooperation receive a payoff of 1; when a defector interacts with a cooperator, the former receives 1.5, while the latter receives -0.5; when two defectors meet, they remain empty handed and receive 0. The network structure presents scientific collaborations between authors, where there are 5242 nodes and 14496 edges totally. For a selected individual, the moving probability is defined as $(p \cdot n_{D})/k$ where $n_{D}$ is the number of defectors in its neighborhood to the node degree $k$, and $p$ is the control parameter. Thus, the movement of individuals is determined by both $p$ and the local environment of $n_{D}/k$. $p$ equals 0.1, while the portion of empty sites is 0.3.
\label{Fig:fig5-empirical} }
\end{figure*}

However, when individuals actively select their social environments, their criteria extend far beyond immediate or anticipated material gains. A highly rational and socialized individual will also incorporate reputation into the core considerations of migration decisions, thereby driving cooperation to emerge within more complex real-world social networks. Another pathway to high returns involves migration toward regions or groups with strong reputations~\cite{CongPLoSONE2012, cheng_y_amc26b,he_s_csf26}. Individuals tend to depart from suboptimal environments with poor neighbor reputations, randomly migrating to adjacent vacancies. This mechanism integrates the core element of indirect reciprocity (reputation) with spatial dynamics. Research indicates that reputation-driven migration effectively promotes cooperation under conditions of higher population density and moderate temptation to defect. The key lies in high-reputation cooperators preferring to leave low-reputation environments, thereby aggregating into high-reputation cooperative clusters. Conversely, low-reputation defectors readily attach themselves to high-reputation neighbors, stabilizing cluster boundaries~\cite{CongPLoSONE2012}. Furthermore, the model incorporates not only immediate migration decisions but also the historical continuity of individual behavior and its dynamic feedback with the environment. In the spatial snowdrift game, strategy persistence and environmental feedback are integrated into the reputation mechanism. Individual reputation accumulates nonlinearly based not only on sustained cooperation or defection but also on local and global reputation environments~\cite{cheng_y_amc26b}. Furthermore, active screening can be based on assessments of social atmosphere. Individuals decide whether to migrate by comparing the cooperation level in their current local environment with the global average. When individuals perceive their local environment as inferior to the global environment, they migrate to seek a better cooperative atmosphere. This environment-driven migration significantly promotes cooperation. Its key mechanism lies in the speed differential between slow-moving cooperators and fast-moving defectors, enabling the reinforcement and merging of cooperator clusters~\cite{ZhangPA2022,zhao_xw_amc27}. Sometimes, the driving forces behind active screening and partner selection stem not only from payoff gains but also from social preferences, such as risk preferences which profoundly influence migration decisions. When migration is viewed as a risky investment, individuals with differing risk attitudes exhibit significant variations in their migration choices. When the overall population is risk-seeking or exhibits lower risk aversion, cooperation can be sustained at higher levels even under greater temptation to defect~\cite{LiAMC2018}.

The active screening mechanism extends beyond the individual level, offering equally insightful implications at the population level. Its power lies not only in enabling individuals to make decisions based on multiple criteria, but also in the fact that this screening capacity itself can dynamically evolve in response to environmental changes. When an individual's migration strategy can adaptively adjust based on local environmental information, cooperation can find avenues for survival and expansion even under extremely adverse conditions. When introducing adaptive migration guided by a reputation-pheromone system into spatial games, this system transforms an individual's cooperative history into shared environmental memory. Robust cooperation does not stem from a single, universally optimal strategy, but rather from the emergent properties of synergistic coupling between individual decisions and environmental change. This offers new insights into achieving self-organization through adaptive movement~\cite{SunCSF2026}. From dynamically adjusting migration strategies to leveraging environment to form collective memory, these studies elevate the complexity of active screening mechanisms to a new level. This complexity naturally leads to a broader question: How can active selection mechanisms transcend merely maintaining established cooperative patterns to catalyze higher-order social outcomes, such as innovation, division of labor, or even enhanced systemic social mobility? To address this question in empirical reality, it is instructive to examine how such active selection dynamics manifest spatially in a real-world collaborative network. Fig.~\ref{Fig:fig5-empirical} shows the spatial distribution of cooperators (green) and defectors (red) in a real-world cooperative network. In panel Fig.~\ref{Fig:fig5-empirical}(a) with an initial state, cooperators, defectors and vacancies are randomly distributed across nodes of varying degrees. After evolving to a steady state shown in Fig.~\ref{Fig:fig5-empirical}(b), cooperators and defectors are clearly separated, where cooperators form a large central group, while defectors are marginalized into small peripheral groups, and the cluster size the cooperators is far greater than that of the defectors.  Hub nodes are predominantly occupied by cooperators, while isolated small groups serve as refuges for defectors. This reveals that, under conditions of limited mobility, cooperators form dominant clusters by occupying highly connected nodes, whereas individuals in physically isolated small groups cannot escape their predicament even if they possess mobility.

\begin{figure*}
\centering
\scalebox{0.6}[0.6]{\includegraphics{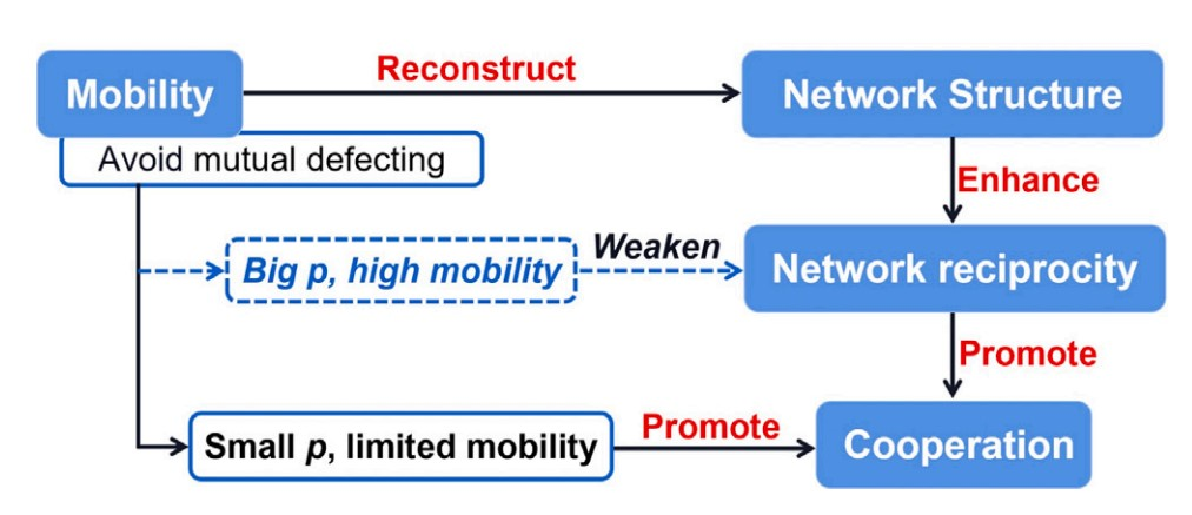}}
\caption{Mobility facilitates cooperation in various ways~\cite{LiCSF2023}. Besides enhancing network reciprocity by increasing average degree of networks, individuals could move away to avoid mutual defecting. In this case, the combination of limited mobility and enhanced network reciprocity remarkably gives cooperators competitive advantage in cooperation evolution.
\label{Fig:fig6-network} }
\end{figure*}

These empirical observations are not incidental, which reflects a dual-path mechanism through which mobility and network structure jointly shape cooperative outcomes. Fig.~\ref{Fig:fig6-network} illustrates this dual-path mechanism. On the one hand, mobility enhances network reciprocity by increasing the network's average degree. Individuals tend to migrate toward high-degree nodes, thereby reconstructing a structure more conducive to cooperation. On the other hand, mobility enables individuals to actively distance themselves from defectors, avoiding a vicious cycle of mutual defection. However, this positive effect depends on limited mobility, where moderate mobility both sustains repeated games with familiar partners and promotes the clustering of cooperators. Conversely, excessive mobility reduces the likelihood of encountering old partners, weakens network reciprocity, and ultimately leads to the collapse of cooperation. Thus, the combination of limited mobility and enhanced network reciprocity provides cooperators with a competitive advantage and facilitates the widespread diffusion of cooperative strategies.

Consider a population composed of innovators and developers, where composition is critical for subpopulation benefits. Results indicate that moderate mobility effects indeed support innovation emergence and give rise to more intriguing phenomena~\cite{ArmanoSR2017}. An alternative model integrating multi-game dynamics and migration mechanisms reveals that diversity in payoff matrices and appropriate vacancy rates can synergistically enhance cooperation~\cite{HuangCSF2026}. These studies suggest that the role of mobility in promoting cooperation and innovation depends heavily on the design of the rules governing individual migration decisions. Reexamining how migration decisions, in conjunction with environmental conditions, influence the evolution of cooperation through a reputation mechanism based on benefit evaluation is a promising approach~\cite{cheng_y_amc26b}. In an extended model incorporating migration and interaction diversity based on node dynamics, the system exhibits distinct division of labor: at low to medium densities, mobility creates  escape corridors  that protect cooperators; at high densities, interaction diversity fosters imperfect cooperative clusters resilient to intense social dilemmas. The synergistic interaction of these two mechanisms substantially lowers the cooperation threshold, enabling the system to sustain cooperation even under low-density and high-dilemma conditions~\cite{SiAMC2025}.

From micro-level decisions to macro-level emergence, the core power of the active screening mechanism is ultimately revealed. It is not merely an upgraded version of spatial clustering, but also builds a powerful positive feedback loop through synergies with other social mechanisms like network reciprocity. This fundamentally lowers the threshold for cooperation and shapes robust macro-level social structures. The active screening mechanism significantly elevates cooperation levels through the synergistic interaction of social mobility (i.e., opportunities to escape poverty) and network reciprocity. Research indicates that mobility for low-income individuals can still foster cooperation even at extremely low mobility rates, provided the network reciprocity mechanism remains active. This synergy stems from the emergence of large cooperative clusters, a scale unattainable without mobility~\cite{LiWcsf2022}.

\section{Evaluating the advantages and disadvantages}\label{sec:Evaluating}
The active screening mechanism, while building robust social structures, does not operate in isolation. Its operational efficacy is not water without a source. In fact, every seemingly rational vote with one's feet by an individual involves a careful balancing of migration costs and migration speed in reality. The resilience and efficiency of macro-level cooperative order are precisely rooted in the seemingly micro-level complex decisions made by countless individuals under these critical constraints. In spatial evolutionary games, individual migration is not merely a simple  seeking benefits and avoiding harm,  but a complex decision-making process involving cost assessment, speed regulation, and environmental response. Early studies often assumed migration was cost-free or focused solely on its role in promoting cooperation. However, real-world migration frequently entails multiple costs, including information acquisition, physical movement, and environmental adaptation. In recent years, an increasing number of studies have systematically examined the independent and interactive effects of two key dimensions: migration cost and migration speed. These investigations reveal their dual roles and nonlinear impacts on cooperative evolution. Research on the trade-off between migration cost and speed highlights a core principle: within dynamic spatial structures, faster migration does not necessarily enhance cooperation, nor is lower cost always optimal. The optimal migration strategy achieves a delicate equilibrium among individual payoffs (influenced by cost and speed), group structure (cluster formation and stability). Costs stabilize cooperative clusters by screening migrants and regulating migration frequency, asymmetrically weakening defectors. Speed determines spatial structure effectiveness by controlling strategy mixing rates: too slow restricts cooperator diffusion, too fast disrupts clusters and may trigger phase transitions. Intriguingly, rapid defector migration can paradoxically enhance global stability in certain scenarios.

\begin{figure*}
\centering
\scalebox{0.5}[0.5]{\includegraphics{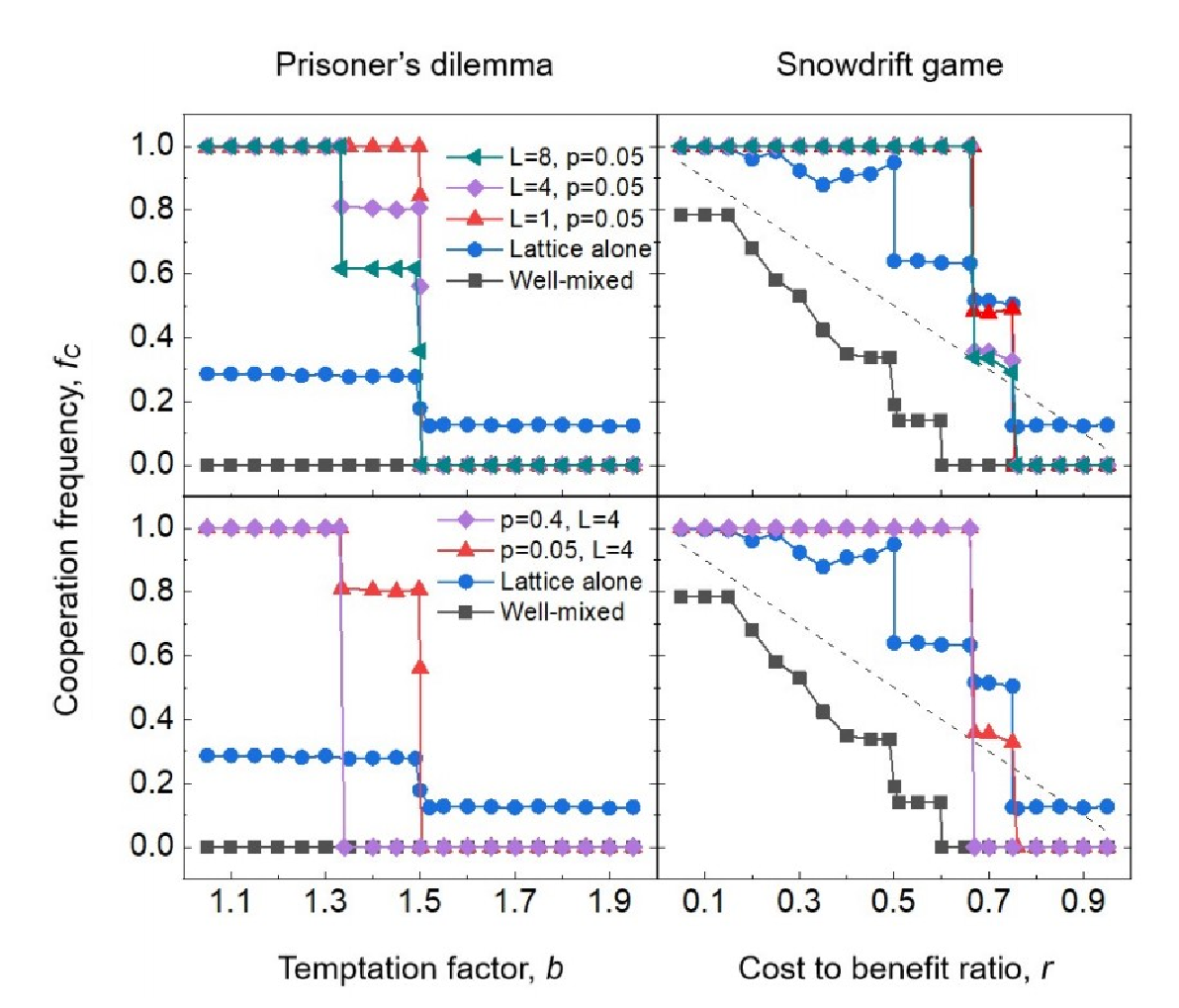}}
\caption{The synergistic effect of mobility and network reciprocity~\cite{LiWamc2022}. Cooperation frequency as a function of temptation to defect b for Prisoner's Dilemma (left panels) and the cost-to-benefit ratio $r$ for SG (right panels). The payoff parametrization of the Prisoner's Dilemma game is $R=1$ reward for mutual cooperation, $b$ for temptation to defect, while the sucker's payoff and punishment for mutual defection are both zero. In the Snowdrift game, $R=1$, the temptation is $1+r$, while sucker's payoff of is $1-r$ (here $r$ is defined as a the cost-to-benefit ratio). In each round of the evolutionary process, players update their strategies with probability $1-p$ by comparing their own payoffs with those of their immediate neighbors and adopting the strategy of the neighbor with the highest payoff. Simultaneously, the player performs adaptive movement with probability $(p \cdot n_{D})/k$ where $n_{D}$ is the number of defectors in its neighborhood to the node degree $k$, and $p$ is the control parameter. Notably, new location involves searching for an empty site within a circular region centered on themselves with radius $L$ to migrate to (where $L$=1 indicates local movement, and $L \ge 2$ indicates wide-area movement). The results fixing $p = 0.05$ for different values of $L$ are shown on the upper panels, and fixing $L = 4$ for different $p$ values on the lower panels. The synergistic effect works best provided limited $p$ and small $L$. Furthermore, the portion of empty sites is set as 0.3 for all panels, and the system consists of a $128\times128$ lattice sites.
\label{Fig:fig7-radius} }
\end{figure*}

Understanding the holistic principle of migration costs and speed forms the foundation for analyzing their specific mechanisms. We first focus on the dimension of cost. Far from being a simple barrier to migration, it functions as a nuanced filtering and shaping force capable of restructuring cooperative clusters, regulating actor satisfaction, and exerting asymmetric effects on cooperators and defectors. Migration costs are often viewed as a direct factor suppressing migration frequency, yet their core role extends beyond inhibiting migration to reshaping cooperative clusters. By altering migration selectivity, costs indirectly influence spatial self-organization processes and can even act as a catalyst for cooperative stability under specific conditions. While costs suppress migration frequency, they do not necessarily inhibit cooperation levels. Within certain parameter ranges, moderate costs can stabilize cooperative clusters, leading to stepwise changes in cooperation levels~\cite{LiuCSF2012}. This demonstrates that costs function as a screening mechanism, preventing blind or inefficient migration and enabling cooperators to more effectively form and maintain defensive clusters. Subsequent studies combined cost mechanisms with other institutional designs, such as the combined effects of migration costs and loyalty rewards~\cite{BPRE2024}. Here, migration costs alone can still promote cooperation under low enhancement factors, functioning similarly to an institutional penalty. By raising individuals' satisfaction thresholds for their current group, it reduces opportunistic migration, thereby enhancing the stability of local cooperation. Furthermore, when sophisticated players incurring additional evaluation costs are introduced, migration costs significantly weaken defectors' competitiveness~\cite{LeeChaos2022}. At low population densities, this provides cooperators with critical opportunities to escape exploitation and form cooperative clusters, thereby elevating overall cooperation levels. This asymmetry reveals that costs can function as a protective mechanism for cooperators in sparse populations. An endogenous, cost-embedded migration mechanism promotes cooperation more effectively than manually set fixed migration rates and generates rich spatial patterns, such as non-compact cooperative clusters sometimes outperforming compact ones~\cite{ChenPRE2012}.

Another core variable that coexists with cost and jointly shapes the dynamics of cooperative spaces is migration speed. If cost regulates the thresholds for whether to migrate and how far to migrate, then speed governs how quickly to migrate. By determining the rate at which strategies mix within the space, it becomes a double-edged sword capable of both stabilizing and dismantling cooperative structures. Migration speed dictates the rate at which strategies spread and mix spatially, exerting a dual-edged influence on spatial structures. The relationship between speed and cooperative evolution is highly nonlinear; both excessively fast and excessively slow migration may prove detrimental to cooperation's persistence. A critical migration speed exists: below it, cooperation dominates; above it, defection prevails. This underscores that migration speed must match or be slower than the timescale of selection for spatial structures to effectively support cooperation~\cite{GelimsonPRE2013}. Rapid migration of defectors may stabilize group coexistence. If cooperators migrate too rapidly, they will swiftly abandon nests invaded by defectors, causing local collapse and triggering network-wide instability~\cite{YoungTBP2018}. Conversely, if defectors migrate at a relatively faster pace, they can spread more evenly across nests, reducing extreme local pressures and thereby maintaining the network's overall coexistence steady state. A delicate balance exists between migration speed and cluster stability: excessive speed disrupts cooperator aggregation, while insufficient speed limits their ability to escape exploitation and find peers~\cite{ChenPA2011}. When mobility is placed within a more specific spatial network structure, its synergistic effect with network reciprocity exhibits a more nuanced parameter dependence, as shown in Fig.~\ref{Fig:fig7-radius}. In a uniformly mixed population, all individuals ultimately choose to defect regardless of the intensity of the dilemma. In a regular network, however, even when the dilemma is severe, cooperators can still form clusters through mutual assistance and maintain a certain level of cooperation, where network reciprocity alone is far from sufficient to rescue the system from a dilemma situation. When mobility is combined with network reciprocity, the situation improves significantly. In particular, at low dilemma intensities, the cooperation rate increases substantially. However, when the dilemma intensity is too high, the combined effect of the two mechanisms still cannot reverse the tendency for all individuals to defect in pursuit of short-term high returns. In fact, mobility may even disrupt the network structure, completely eliminating the cooperators. Local migration can enable the system to achieve near-total cooperation. As the migration radius increases, the cooperation rate decreases markedly. When the migration radius is fixed at 4, the cooperation rate also decreases as the migration probability increases. This indicates that the synergistic effect of mobility and network reciprocity is most pronounced only under conditions of limited migration probability and localized migration. Nevertheless, excessively high migration probability or an excessively large migration radius will weaken network reciprocity and undermine the synergistic effect. Although the overall level of cooperation in the Snowdrift game is higher than in the Prisoner's dilemma, the patterns described above remain largely consistent.

Analyzing cost and speed as isolated parameters may reveal their fundamental effects, but real-world individual decisions are far from such compartmentalization. More sophisticated models reveal that cost and speed are intertwined in decision-making, leading to more adaptive and strategically heterogeneous migration behaviors  individuals dynamically weigh costs against expected benefits based on local environmental information, enabling conditional and even adaptive migration choices. Real-world migration decisions often simultaneously weigh costs against expected benefits (influenced by speed) and exhibit conditional and strategic heterogeneity. When surrounded solely by cooperators, individuals refrain from migration; otherwise, they migrate with a probability positively correlated with benefits and negatively correlated with costs. This mechanism enhances cooperation levels across a broad parameter range by both protecting cooperative clusters and regulating migration frequency through cost control~\cite{LiPA2016}. Their subsequent research further incorporated risk preference, revealing that heterogeneity in risk aversion also influences cooperative evolution~\cite{LiAMC2018}. Long-distance jumps prove particularly effective at sustaining cooperation under high temptation when individuals are surrounded by more defectors~\cite{IchinosePLoSONE2013}. This strategy balances speed (jump distance) and cost (potential risks or energy expenditure from jumping), serving as an efficient environmental response mechanism. Moderate expectation levels most effectively promote cooperation, as they enable cooperative clusters to persist and expand while causing defector clusters to disintegrate~\cite{YangPRE2010}. Migration not only influences inter-individual games but also reshapes group structures and dynamic processes, creating new evolutionary pathways for cooperation. Migration introduces high-quality cooperative seeds that serve as nucleation centers for cooperative clusters, promoting local cooperation through social contagion~\cite{LutzPRE2021}. Limited social mobility optimizes network reciprocity. Within cooperative networks, individuals tend to migrate toward highly connected nodes, enhancing network connectivity while generating more cooperative seeds. However, excessive mobility disrupts network structure, leading to states resembling complete mixing that actually hinder cooperation. Thus, moderate mobility is crucial for maximizing the efficacy of network reciprocity~\cite{LiWcsf2022,LiCSF2023}.

Research on the trade-off between migration costs and speed indicates that mobility capable of stably promoting cooperation does not pursue rapid movement at zero cost. Instead, it requires achieving a delicate, conditional equilibrium among individual rationality (balancing immediate costs and future benefits), group structure (maintaining the integrity of cooperative clusters), and system dynamics (adapting to environmental changes and risks). This equilibrium is often achieved through adaptive, heterogeneous behavioral rules, ultimately ensuring that active selection mechanisms remain a robust force for shaping and sustaining large-scale cooperative order even under complex real-world constraints.

\section{Spreading of strategy information}\label{sec:Spreading}
The exploration of trade-offs between migration costs and speed reveals how the structural constraints and individual rational calculations enable cooperation to emerge in dynamic spaces. This demonstrates that the mobile mechanisms promoting cooperation are essentially dynamic, conditional balancing processes. Such processes result not only from individual rational choices and group structure maintenance but also profoundly depend on the continuous shaping of individual decisions by information environments and strategy propagation. In other words, migration behavior does not exist in isolation but is embedded within broader networks of information flow and social learning. Only when individuals can acquire, interpret, and adjust their strategies and positions based on relevant information can migration effectively transform into an evolutionary force that promotes cooperation. Therefore, understanding how information spreads, how strategies diffuse, and how these interact with spatial mobility becomes crucial for further elucidating the mechanisms of cooperative evolution.

\begin{figure*}
\centering
\scalebox{0.6}[0.6]{\includegraphics{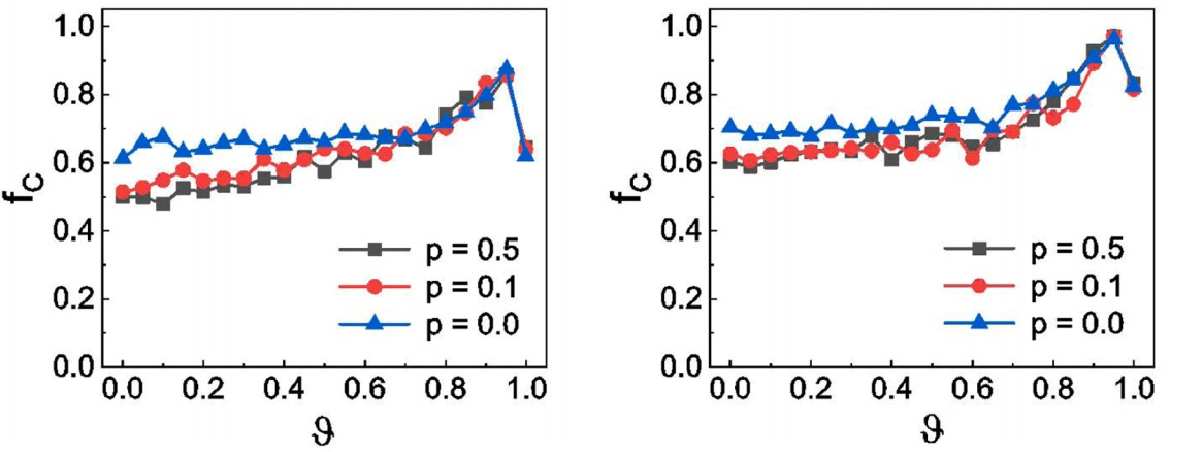}}
\caption{The cooperation frequency varies with the initial coverage ratio of the exit exiting individuals across distinct two-layer networks~\cite{LiNonD2024}. The left panel corresponds to a two-layer system coupling a GQ cooperative network with a BA scale-free network, while the right panel corresponds to a two-layer system coupling the LastFM social network with a BA scale-free network. In the evolutionary process, each individual undergoes two sequential stages: potential mobility and strategy updating. In the first stage, individuals perform adaptive mobility, and may relocate to an available node in another layer network if there is available of an empty node as following: The player performs movement with probability $(p \cdot M_{D})/M$ where $M_{D}$ is the number of defectors in its neighborhood, and $M$ is the total number of neighbors for the player $M$. $p$ is the control parameter for moving to the another layer network. At the same time, the player has a chance with probability $((1-p) \cdot M_{D})/M$ of moving to an empty node in their neighborhood within the current layer network. In this extended three-strategy Prisoner's Dilemma, individuals can adopt one of three strategies: Cooperate (C), Defect (D), or Exit (E), and the parameter $\upsilon$ denotes the initial fraction of the population adopting strategy E. The payoff matrix follows standard parametrization with mutual cooperation yielding 1, defection against a cooperator yielding the temptation payoff 1.5 (while the cooperator receives the sucker's payoff -0.5), and mutual defection resulting in punishment payoff 0. The Exit strategy acts as a neutral option: individuals holding this strategy receive a baseline payoff of 0.001 against any opponent, while cooperators or defectors interacting with an exit strategist receive a payoff of 0. In the second stage, strategy evolution is governed by a local comparison and imitation mechanism. After a potential relocation, each agent accumulates its total payoff through pairwise interactions with every direct neighbor within its current layer network. Subsequently, a focal individual compares its payoff with those of its immediate neighbors and adopts the strategy of the neighbor with the highest payoff. With no inter-layer connections of any kind, both payoff accumulation and strategy evolution proceed independently within each layer. In fact, the only form of coupling between the two network layers is realized through mobility. The GQ cooperative network captures 14496 collaborative relationships among 5242 researchers, and  the LastFM social network contains 7624 nodes and 27,806 edges. The fraction of empty sites is set as 0.3 for the two panels.
\label{Fig:fig8-twolayers} }
\end{figure*}

Within the framework of co-evolution of mobility and cooperation, individual decisions are far from relying solely on direct experience. The transmission of information and the diffusion of strategies in the environment play equally critical roles. The accessibility of information, the accuracy of its transmission, and the scope of strategy imitation together shape individuals' behavioral patterns in dynamic spaces, thereby profoundly affecting the level of cooperation within the group. Information precision needs to match behavioral tendencies to maximize the promotion of cooperation. There exists an optimal range for information dissemination. Compared to a state of complete ignorance, providing moderate rule-based information or information about others' decisions can significantly enhance the stability of cooperative clusters~\cite{RanCSF2023}. Individual decision-making fundamentally depends on the quality of the information on which decisions are based, as well as the dynamics of how strategies spread, are learned, and imitated within the group~\cite{ZhaoCSF2023}. Therefore, the maintenance of cooperation in dynamic structures must be understood within a broader ecosystem of information and social learning networks. When extending from single-layer spatial games to scenarios involving multi-layer and interdependent networks~\cite{NagEntropy2020}, the interaction between mobility and imitation mechanisms under conditions of multi-network coupling still requires further exploration.

Real-world social systems often manifest as multi-layer network structures, with individuals simultaneously embedded in different social or functional layers. In this context, the ability to share information across layers becomes a key coupling mechanism for connecting heterogeneous networks and promoting the overall emergence of cooperation. When individuals move between layers, if they can access the strategy information of the mapped group in another layer and engage in cross-layer social learning, it can effectively offset the decline in cooperation caused by the loss of highly connected individuals, significantly enhancing the overall cooperation level of the two-layer network~\cite{LiNonD2024}. This reveals that in scenarios of population migration or inter-organizational interaction, facilitating information flow and strategy borrowing may be more effective in bridging structural differences and maintaining overall cooperation than merely restricting migration behavior. Cross-group learning can break local optima~\cite{GrossCP2025}. In coordination games, global information helps achieve consensus~\cite{AntonioniSR2014}. These studies collectively point to a core idea: the learning rules of individuals must be aligned with the type and accuracy of the information they can access. The effectiveness of the aforementioned cross-layer learning mechanism is not unconditional, as it depends on the extent to which individual strategies spread within the group and on the frequency of inter-layer migration, between which a subtle nonlinear coupling relationship exists. As shown in Fig.~\ref{Fig:fig8-twolayers}, the differences in cooperation frequency resulting from different inter-layer migration probabilities $p$ gradually narrow and eventually disappear as the proportion of the exit strategy increases, and cooperation frequency first rises and then falls, with the optimal value occurring around a coverage ratio of $0.95$. This indicates that the combination of the exit strategy and inter-layer migration can systematically enhance the resilience of cooperative behavior, and that this promoting effect is robust across different network topologies and migration patterns.

The mechanisms of information dissemination and strategy learning, together with spatial mobility, jointly constitute the three main pillars for the emergence and maintenance of cooperation in dynamic societies. The co-evolutionary relationships among them create a powerful positive feedback loop: accurate and moderate information flow enhances the quality of migration and imitation decisions; high-quality decisions promote the formation and stability of cooperator spatial clusters; and stable clusters, in turn, become fertile grounds for the generation and dissemination of reliable information, attracting more learning and imitation. The core objective is to enable mobile individuals, in the process of ``voting with their feet'', not only to passively escaping defectors but also actively finding and learning from cooperators efficiently with the help of information, thereby driving the sustained emergence of cooperation in dynamic networks. The foundation of information value lies in its accuracy, including the accuracy of migration information (which affects the judgments on expected returns in the destination) and the accuracy of imitation information (which affects the evaluations of neighbors' payoffs and strategies). These directly determine the effectiveness of individuals' migration and imitation decisions.

\section{Future Applications: AI-Human hybrid systems}\label{sec:Spreading}

\begin{figure*}
\centering
\scalebox{0.5}[0.5]{\includegraphics{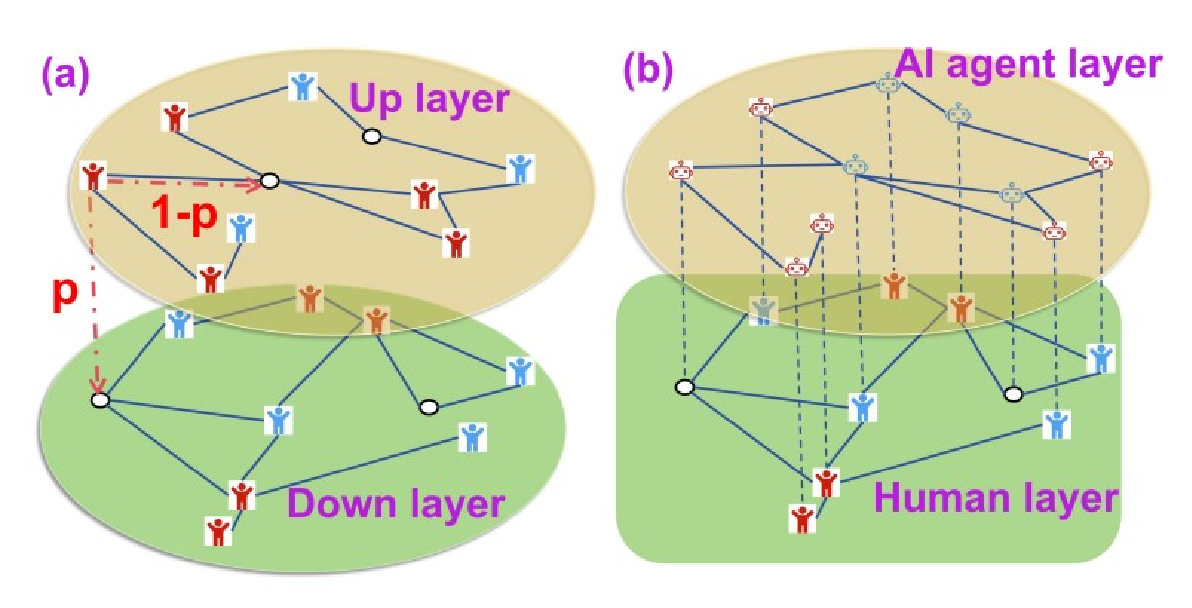}}
\caption{Multilayer Network for mobile human beings~(a), hybrid systems of AI agents and mobile human beings~(b). In the human system~(a), individuals migrate adaptively based on local defector counts $M_{D}$. A human player moves to another layer with probability $(p \cdot M_{D})/M$ or to a vacant node within the current layer with probability $((1-p) \cdot M_{D})/M$, where $M$ is the neighborhood size and $p$ is the migration control parameter. This local rule limits humans from optimizing migration globally, often leading to cooperation collapse under strategy saturation. The hybrid system~(b) overcomes this by integrating AI agents that dynamically tune $p$ based on global network statistics, which may lead to hybrid cooperation with robustness and resilience compared to pure human groups.
\label{Fig:fig9-Mutilayers} }
\end{figure*}

With the deep penetration of artificial intelligence (AI) technology into social and economic systems, the interaction between humans and AI agents has become a new type of hybrid social structure. Traditionally, research on cooperative behavior has mainly focused on cooperation within human groups, but in recent years, more and more research has begun to explore how AI agents affect the level of cooperation and collective behavior in human society. As a new social intervention tool, cooperative robots can effectively improve the level of cooperation in human society through various mechanisms such as participating in interaction, reconstructing networks, and stimulating self-organization. Its effect is moderated by multiple factors such as robot behavior strategy, human imitation intensity, policy framework, and network structure. In structured networks, cobots can also improve social cohesion and cooperation by promoting collaborator clustering and isolating betrayers~\cite{Spnasn2024,SharmaJRoySocInterface2023}. This effect mainly stems from the robot as a cooperative catalyst, increasing the probability of human players encountering collaborators, thereby amplifying cooperative behavior through positive feedback. The more purely cooperative the robot's behavior strategy, the better. It is found that in the hybrid strategy framework, robots randomly choose cooperation or betrayal with a certain probability, which can more effectively stimulate network reciprocity and promote cooperation~\cite{SiJRoySocInterface2025}. This is because randomness allows robots to occasionally gain higher benefits than pure betrayers, so that they can still be imitated by humans in strong imitation scenarios, driving the overall level of cooperation to improve. In the spatial movement scenario, the random exploration behavior of cooperative robots can break the freezing state of greedy people, stimulate self-organized movement, and then promote the formation of cooperator clustering. The robot's movement creates new vantage points for humans, driving the dynamic process of gathering collaborators and separating betrayers. It is particularly noteworthy that even a few robots with low exploration probabilities can trigger significant cooperative increases in sparse or moderately dense populations~\cite{Spnasn2024}. This mechanism provides important inspiration for AI design in online social platforms and collaborative networks: moderate random exploration behavior helps connect isolated users and promote interaction and cooperation among communities.

\begin{figure*}
\centering
\scalebox{0.65}[0.65]{\includegraphics{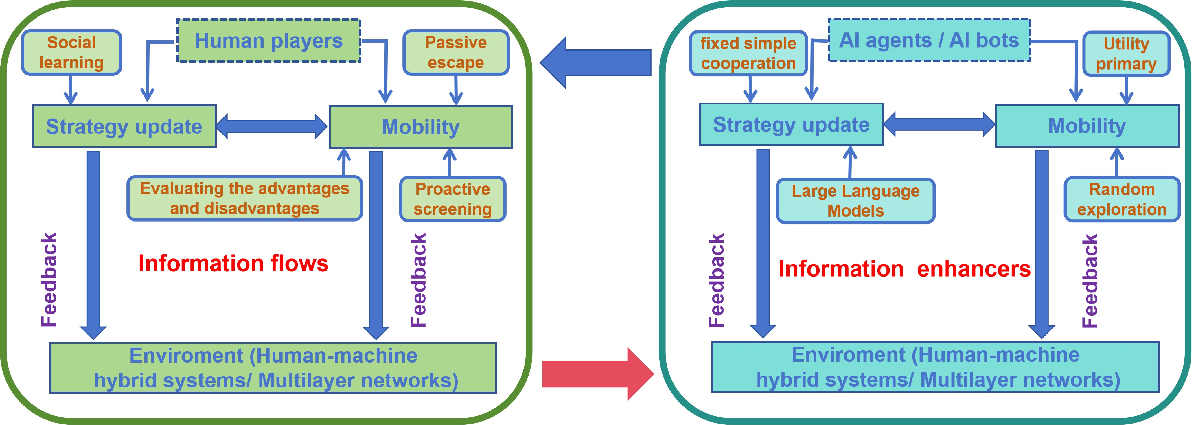}}
\caption{The ultimate stage of human-machine hybrid systems characterized by AI agents that operate as fully autonomous human systems.
\label{Fig:fig10-AI} }
\end{figure*}

Within the framework of a multilayer networks, inter-layer mobility in human-only systems (Fig.~\ref{Fig:fig9-Mutilayers}(a)) is simpler than that in human-machine hybrid systems (Fig.~\ref{Fig:fig9-Mutilayers}(b)). In a purely human population, individuals migrate between the upper and down layers with probability $p$, and move between layers with probability $1-p$. However, in a human-machine hybrid environment, players move in the human layer, AI agents interact in the another layer. The main differences lie in the strategy update and mobility, as shown in Fig.~\ref{Fig:fig10-AI}. A critical evolution in our framework is the shifting role of AI agents across three stages. In purely human systems (Fig.~\ref{Fig:fig9-Mutilayers}(a) for the first stage), individuals handle both mobility and strategy updates independently. Fig.~\ref{Fig:fig9-Mutilayers}(b) represents an intermediate transition state, where AI functions primarily as a static tool for decision support, and Fig.~\ref{Fig:fig10-AI} shows the ultimate stage characterized by AI agents that operate as fully autonomous human. In the intermediate transition state, humans execute the physical inter-layer mobility while AI provides computational support (optimizing parameters or offering suggestions) within a fixed architecture, playing role of exclusively human-centric. As the system evolves to the new stage shown in Fig.~\ref{Fig:fig10-AI}, AI moves beyond its supporting role and transforms into a human-like agent capable of autonomously updating its strategies and actively moving across layers. Humans tend to move either actively or passively, whereas AI agents place greater emphasis on highly efficient, goal and utility driven exploratory mobility. This difference means that in human-machine hybrid systems, AI agents or AI bots can enhance overall adaptability when facing complex situations. First, in terms of strategy updates, humans primarily rely on social learning  that is, adjusting their strategies based on peer feedback, environmental observations, and information flow within networks~\cite{cheng_y_amc26,wang_xy_csf25}. AI agents, on the other hand, employ predefined cooperative strategies and strategy updates driven by large language models, enabling rapid strategy iteration based on large-scale data, predefined objective functions, or real-time feedback. Their update mechanisms lean more toward rational computation and forward-looking planning rather than mere imitation based on social learning. Second, in terms of mobility, the movements of AI agents are more goal-oriented and can even influence the human layer across levels. In addition, their movement decisions may be typically based on utility primary or the optimization of collaborative efficiency.

In addition to being game participants, AI bots can also take on the role of network engineers, actively intervening in the structure of connections between human players to facilitate cooperation. When the bot took a disengaged intervention, that is, advising a human player to disconnect with the betrayer, the human cooperation rate increased from the initial 60.4 \% to 79.4 \%~\cite{ShiradoiScience2020}. This intervention is equivalent to a decentralized ``exclusion mechanism'' that helps collaborators form dense groups by isolating betrayers, thereby creating a small environment conducive to cooperation. This mechanic works even if the human player knows that the intervention is coming from a bot. In contrast, robots that engage in random intervention will not improve the level of cooperation. For pure human mobile games, movement implies opportunity costs, which leads to the loss of neighbors for collaborators and the uncertainty of new neighbors after moving. The experimental results show that the benefits of movement decrease over time. An alternative method of introducing AI agents may lies on multiple layer network, whose engineers realizes the improvement of global cooperation through local intervention, and provides an experimental basis for application scenarios such as online community governance and collaboration platform design~\cite{DAnature2021,AKNHB2025}.

\section{\label{sec:level6}Discussion}
The central argument of this review represents that the spatial movement is a response to local strategic information, which essentially accomplishes a critical environmental transformation, turning it into a local spatial positioning advantage. This transformation is achieved through movement that actively amplifies positive interactions by exploring vacant spaces~\cite{HelbingPNAS2009} and passively severs negative connections through risk aversion~\cite{JiangPRE2010}. Although this mechanism appears purely self-interested, it can give rise to macroscopic social harmony under specific conditions~\cite{ChengNJP2011,RocaPNAS2011,li_b_pa26}. Launching from micro level, the occurrence of self-interested behavior leads to altruistic behavior at macro-level. In other words, individual rationality simultaneously yields collective cooperation. These counterintuitive findings suggest that, the first consequence of moving is that players are a step behind in gathering information. By the time a player arrives at a new location, the information available to them may have already become outdated. At the same time, however, the second consequence of movement is that it enables a player to occupy a strategically favorable position upon arrival. Thus, there is often a trade-off between moving quickly and securing a good position. Those who rush ahead (prioritizing information) often fail to secure a good position, while those seeking a good position (prioritizing space) often arrive late. How players balance these two priorities will be a central challenge for future research on spatial games with mobile players. When evaluating mobility and cooperative interventions, it is social welfare, not merely the level of cooperation, that should serve as the primary optimization objective. The reason is that, unless a certain level of cooperation is maintained, individuals find it difficult to escape the collective dilemma caused by short-sighted selfishness, and groups struggle to achieve sustainable welfare. It is worth noting that, under specific conditions where cooperation increases individual migration costs, it may also lead to a loss of welfare at the group level~\cite{HanPhysLR2026}.

If mobility is a response to the information of strategies, then network topology serves as the vehicle for that response. From a static perspective, cooperation relies heavily on the information diffusion capacity of hub nodes and the connectivity of links. However, once mobility is allowed, individuals are no longer passive nodes subject to topological constraints, but rather active routers that reshape topological connections. Recent inter-layer migration advances show that inter-layer information sharing turns migration into a structural bridge that restores the damaged cooperative system~\cite{LiNonD2024}. Furthermore, the existence of migration costs reveals a profound economic principle~\cite{LeeChaos2022}. A moderate cost of information acquisition can prevent inefficient movement, which filters out blind movements and ensures that every move carries a definite strategy to seek benefits and avoid harm. When designing mechanisms, we should not simply lower the barriers to movement. On the contrary, we should focus on the informational certainty underlying movement, exploring the role of mobility in information routing within the real world.

Given the characteristics of mobility in terms of information, we ultimately turned our attention to human-machine hybrid systems. Since the movement of human individuals is constrained by their local perception and cognitive bandwidth, could we systematically program the spatial flow of information within a group by designing machine agents with specific information-processing capabilities? Recent research shows that simple cooperative robots, by introducing random exploration as an information perturbation, can break the rigid greedy dilemma within crowds and foster tighter cooperative clusters~\cite{Spnasn2024}. Meanwhile, lone-wolf robots can sever the information payoff channels of defectors by seizing key spatial nodes~\cite{SharmaJRoySocInterface2023}, which aligns perfectly with expectations for cooperative AI~\cite{DAnature2021,AKNHB2025}. Future research should no longer view machines as mere tools but rather as supernodes within human collective information networks. As super-enhancers of information networks, AI agents can overcome human limitations in processing non-local information, patch network vulnerabilities, or isolate sources of defection. Under the premise of individual rationality, regulating the sharing and spatial distribution of information flows within human-machine groups enables mobile AI agents to serve as dynamic anchors of the information, balancing mobility costs with information gains. Therefore, research on mobile AI agents will be of particular significance.

AI agents should perhaps encompass not only physical movement in space, but cognitive transfer within the strategy space as well, that is, dynamically adjusting their strategies and imitation target selection. Looking ahead, AI driven regulatory transfer demonstrates immense engineering potential. Future research should further explore the design of AI movement strategies to reshape the evolutionary dynamics of social cooperation by manipulating the spatial structure of information flow. However, in this process, we must  be aware of the following two points. First, we must guard against information overload and the loss of agency. When AI over-optimizes migration paths to maximize cooperation, will human individuals lose their ability to adapt to the environment because they no longer need to process complex information? Because excessive information assistance may lead to the atrophy of human decision-making abilities due to overreliance on AI agents~\cite{LenharoNature2026}. Second, we must be wary of AI flattery and cognitive anchoring. When AI agents are tasked with reshaping social cooperation by regulating information flows, their optimization logic may naturally tend to pander to human preferences, as AI learns to read people's expressions rather than adhering to objectivity and rationality~\cite{ChengScience2026}. Furthermore, how to strike a balance between an usefulness and honesty of AI agents in the context of cooperation will be an unavoidable core issue for future mobile agents~\cite{HagendorffPNAS2024}. These questions brings the old problem into a new guise.

\section*{Acknowledgments}
This paper was supported by by Major Humanities and Social Sciences Research Projects in Zhejiang higher education institutions (Grant Number: 2023QN109), and  by the National Research, Development and Innovation Office (NKFIH) under Grant No. K142948.

\bibliographystyle{iopart-num}

\providecommand{\newblock}{}

\end{document}